\documentclass[aps,prc]{revtex4}
\usepackage{amssymb,amsmath,epsfig, psfrag, ulem, color,graphicx}
\newcommand{\rf}[1]{(\ref{#1})}
\newcommand{\ba}{\begin{align}}
\newcommand{\ea}{\end{align}}
\newcommand{\lam}{\Lambda}
\newcommand{\hate}{\mathbf{\hat e}}

\newcommand{\half}{\frac{1}{2}}

\newcommand{\be}{\begin{equation}}  
\newcommand{\ee}{\end{equation}}  
\newcommand{\bea}{\begin{eqnarray}}  
\newcommand{\eea}{\end{eqnarray}}  
\newcommand{\bmu}{\begin{multline}}
\newcommand{\emu}{\end{multline}}
\newcommand{\bean}{\begin{eqnarray*}}  
\newcommand{\eean}{\end{eqnarray*}}

\newcommand{\gapproxeq}{\lower  
.7ex\hbox{$\;\stackrel{\textstyle >}{\sim}\;$}}  
\newcommand{\lapproxeq}{\lower  
.7ex\hbox{$\;\stackrel{\textstyle <}{\sim}\;$}}

\newcommand{\bc}{\begin{center}}  
\newcommand{\ec}{\end{center}}  
\newcommand{\btab}{\begin{tabular}}  
\newcommand{\etab}{\end{tabular}}

\newcommand{\cleb}[6]{\langle{#1}  {#2},{#3}  {#4}|{#5}  {#6}\rangle}
\newcommand{\qq}{q\overline q}
\newcommand{\cc}{c\overline c}
\newcommand{\bb}{b\overline b}

\newcommand{\bk}[2]{\langle{#1}|{#2}\rangle}
\newcommand{\ket}[1]{|           {#1}           \rangle}
\newcommand{\bra}[1]{\langle           {#1}           |}   
\newcommand{\tpn}{^3\mathrm{P}_0}
\newcommand{\tso}{^3\mathrm{S}_1}  
\newcommand{\grad}{\mathbf{\nabla}}
\newcommand{\sixj}[6]{ \left\{\begin{array}{ccc}  {#1} & {#2} & {#3} \\    {#4} & {#5} & {#6} \end{array}   \right\} }
\newcommand{\ninej}[9]{ \left\{\begin{array}{ccc}  {#1} & {#2} & {#3} \\    {#4} & {#5} & {#6}\\    {#7} & {#8} & {#9}  \end{array}   \right\} }
\newcommand{\rme}[3]{\bra{#1}|{#2}|\ket{#3}}
\newcommand{\me}[3]{\bra{#1}{#2}\ket{#3}}
\newcommand{\ot}{\otimes}
\newcommand{\st}[1]{\Pi_{#1}}
\newcommand{\vecr}{\mathbf{r}}
\newcommand{\vecrho}{\boldsymbol{\rho}}
\newcommand{\vecR}{\mathbf{R}}
\newcommand{\vecl}{\mathbf{l}}
\newcommand{\vecls}{\mathbf{l}_s}

\newcommand{\veca}{\mathbf{a}}
\newcommand{\vecp}{\mathbf{p}}
\newcommand{\vecy}{\mathbf{y}}
\newcommand{\hatr}{\mathbf{\hat r}}
\newcommand{\hatrho}{\boldsymbol{\hat \rho}}
\newcommand{\hatR}{\mathbf{\hat R}}
\newcommand{\ud}{\mathrm d}

\newcommand{\uS}{\textrm{S}}
\newcommand{\uP}{\textrm{P}}
\newcommand{\uD}{\textrm{D}}
\newcommand{\uF}{\textrm{F}}

\newcommand{\ssn}{^1\textrm{S}_0}
\newcommand{\tdo}{^3\textrm{D}_1}
\newcommand{\lambar}{\overline\lam}
\newcommand{\epem}{e^+e^-}
\def\bX{\mathbf{X}}

\begin{document}

\title{\bf Strong decay amplitudes using the Talmi-Moshinsky transformation and the suppressed production of the $1^{-+}$ hybrid}

\author{{\em }
T. J. Burns\footnote{e-mail: burns@thphys.ox.ac.uk}}
\affiliation{ Rudolf Peierls Centre for Theoretical Physics,
University of Oxford, Keble Rd., Oxford, OX1 3NP, United Kingdom}

\begin{abstract}
The Talmi-Moshinsky transformation is applied to the strong decay and production of mesons. Amplitudes for conventional mesons are obtained in a simple way and are generalised to encompass the production and decay of hybrid states. The selection rule disfavouring hybrid decay into pairs of identical S-wave mesons has a natural explanation and a new class of selection rules is uncovered with implications for production of the exotic $1^{-+}$ hybrid in $\epem$ experiments, charmonia decays and the lattice. 
\end{abstract}

\maketitle
%%%%%%%%%%%%%%%%%%%%%%%%%%%%%%%%%%%%%%%%%%%%%%%%%%%%%%%%%%%%%%%%%%%%%%%%%%%%%%%%%%%%%%%%%%%%%%%%%%%%%%

\section{Introduction}

The flux tube model sets a scale for hybrid decays relative to those of conventional mesons \cite{ip,ikp,cp95}, a process which the lattice is now able to calculate \cite{mm06}. The agreement between the two is good in a simple version of the model with equal wavefunction widths and a simplified string overlap term \cite{ftlattice}. A decay formalism is developed here which works within the same level of approximation and treats the production and decay of states with arbitrary gluonic excitation. A technique borrowed from nuclear physics -- the Talmi-Moshinksy transformation -- provides a simple framework for the calculation and uncovers new selection rules sensitive only to the gluonic angular momentum of the initial state. Using the effective quark-flux tube wavefunctions developed elsewhere \cite{tjbunpublished} decays involving hybrid states are a straightforward extension of those involving only conventional mesons. 

Within this formalism the amplitude for the decay of any meson into any other pair of mesons is obtained, be they orbitally, radially or gluonically excited (hybrid) states. Previous results for conventional  and hybrid  meson decays are derived in a  simple way and  new results obtained for hybrids in the final state. The amplitudes are left in rather general form: simpler polynomial forms will be collated elsewhere and discussed with reference to the phenomenology of hybrids in experiment and on the lattice \cite{ccdecay}. The selection rule against the decay of a hybrid meson into pairs of identical S-wave states can be generalised using the symmetry of the Talmi-Moshinsky transformation, giving several new results. A new selection rule forbids the decay of any meson with the gluon field in its ground state to the $1^{-+}$ hybrid and a 1S meson in the limit of equal width outgoing meson wavefunctions.  Within model assumptions developed elsewhere this rule may help discriminate interpretations of the  $X(3940)$ and $Y(4260)$ states, and modes such as 
$
\psi\to \pi_1\rho
$
and $\chi\to\pi_1\pi$
are forbidden.

The Talmi-Moshinksy transformation and its relevance to quark-pair creation processes is discussed in section \ref{talmiandqpc} and applied to the decay of a  conventional meson to any other pair of conventional mesons in section \ref{sss}. Section \ref{effwavefns} introduces effective wavefunctions for mesons with flux tubes which and these are applied to the  decay and production of hybrids in sections \ref{lss} and \ref{lll}. In section \ref{suppressed} some phenomenology of hybrid production in $\epem$ experiments is discussed.

\section{Quark-pair creation and the Talmi-Moshinsky transformation}
\label{talmiandqpc}

Central forces in nuclear physics require matrix elements of the form
\be
\me{(n_1l_1\ot n_2l_2)_{l_{12}m_{12}}}{V(|\vecr_1-\vecr_2|)}{(n_1'l_1'\ot n_2'l_2')_{l_{12}'m_{12}'}}
\ee
where the coupled wavefunctions in the nucleon coordinates $\vecr_1,\vecr_2$ are
\be
\bk{\vecr_1,\vecr_2}{(n_1l_1\ot n_2l_2)_{l_{12}m_{12}}}=
\sum_{m_1 m_2}
\bk{\vecr_1}{n_1l_1m_1}\bk{\vecr_2}{n_2l_2m_2}
\cleb{l_1}{m_1}{l_2}{m_2}{l_{12}}{m_{12}}.
\ee
In the basis of equal width harmonic oscillator wavefunctions there is a convenient procedure for the evaluation of such matrix elements. The wavefuctions in $\bra{\vecr_1,\vecr_2}$ are expanded in a finite sum of wavefunctions in vectors $\bra{\vecR,\vecr}$ proportional to the relative and centre of mass coordinates of the nucleons,
\be
\vecR=\frac{\vecr_1-\vecr_2}{\sqrt 2},\qquad \vecr=\frac{\vecr_1+\vecr_2}{\sqrt 2},
\label{jacobi}
\ee
where to match onto later notation the conventions of upper and lower case to denote the relative and centre of mass coordinates are the opposite of those normally used. Denoting by upper case letters the quantum numbers in $\vecR$ and lower case letters the quantum numbers in $\vecr$, the expansion gives
\begin{multline}
\me{(n_1l_1\ot n_2l_2)_{l_{12}m_{12}}}{V(|\vecr_1-\vecr_2|)}{(n_1'l_1'\ot n_2'l_2')_{l_{12}'m_{12}'}}=
\\
\sum_{NLnl}\sum_{N'L'n'l'}
\bk{(n_1l_1\ot n_2l_2)_{l_{12}      }}{(NL \ot nl)_{l_{12}      }}
\me{(NL \ot nl)_{l_{12} m_{12}     }}{V(\sqrt 2 R)}{(N'L' \ot n'l')_{l_{12}' m_{12}'     }}
\\
\bk{(N'L' \ot n'l')_{l_{12}'       }}{(n_1'l_1'\ot n_2'l_2')_{l_{12}'     }}.
\label{tensorexpansion}
\end{multline}
The centre of mass co-ordinates integrate out entirely   leaving only a simple one dimensional radial integral,
\be
\me{(NL \ot nl)_{l_{12} m_{12}     }}{V(\sqrt 2 R)}{(N'L' \ot n'l')_{l_{12}' m_{12}'     }}
=
\rme{NL}{V(\sqrt 2 R)}{N'L'}\delta_{nn'}\delta_{ll'}\delta_{l_{12}l_{12}'}\delta_{m_{12}m_{12}}'\delta_{LL'}.
\ee
The coefficients in the coordinate transform,
\be
\bk{(n_1l_1\ot n_2l_2)_{l_{12}      }}{(NL \ot nl)_{l_{12}      }}
=\frac{1}{2l_{12}+1}\sum_{m_{12}}
\int\ud^3\vecr_1\ud^3\vecr_2
\bk{(n_1l_1\ot n_2l_2)_{l_{12}m_{12}      }}{\vecr_1,\vecr_2}
\bk{\frac{\vecr_1-\vecr_2}{\sqrt 2},\frac{\vecr_1+\vecr_2}{\sqrt 2}}{(NL \ot nl)_{l_{12} m_{12}   }},
\label{transformationbrackets}
\ee
 are the Talmi-Moshinsky brackets \cite{talmi,mosh} whose values are well known \cite{buck}. The essential idea is that the harmonic oscillator Hamiltonian can be written equivalently in either pair of Jacobi coordinates $(\vecr_1,\vecr_2)$ or $(\vecr,\vecR)$: each construction shares common eigenvalues in ${\mathbf{l}_{12}}^2$, $m_{12}$ and $H$.  The latter demands that the number of energy quanta is conserved,
\be
2n_1+l_1+2n_2+l_2=2N+L+2n+l,
\label{energyconservation}
\ee
so the summations in \rf{tensorexpansion} are always finite. 

\begin{figure}
%qpc.tex
\setlength{\unitlength}{0.00023300in}%
\begingroup\makeatletter\ifx\SetFigFont\undefined
% extract first six characters in \fmtname
\def\x#1#2#3#4#5#6#7\relax{\def\x{#1#2#3#4#5#6}}%
\expandafter\x\fmtname xxxxxx\relax \def\y{splain}%
\ifx\x\y   % LaTeX or SliTeX?
\gdef\SetFigFont#1#2#3{%
  \ifnum #1<17\tiny\else \ifnum #1<20\small\else
  \ifnum #1<24\normalsize\else \ifnum #1<29\large\else
  \ifnum #1<34\Large\else \ifnum #1<41\LARGE\else
     \huge\fi\fi\fi\fi\fi\fi
  \csname #3\endcsname}%
\else
\gdef\SetFigFont#1#2#3{\begingroup
  \count@#1\relax \ifnum 25<\count@\count@25\fi
  \def\x{\endgroup\@setsize\SetFigFont{#2pt}}%
  \expandafter\x
    \csname \romannumeral\the\count@ pt\expandafter\endcsname
    \csname @\romannumeral\the\count@ pt\endcsname
  \csname #3\endcsname}%
\fi
\fi\endgroup
\begin{picture}(5788,4588)(1757,-12705)
\thicklines
\put(5401,-9661){\vector( 1, 1){1050}}
\put(6676,-12436){\vector(-1, 1){975}}
\put(5701,-10411){\vector( 1,-1){975}}
\put(7501,-8611){\vector(-1,-1){1050}}
\put(5101,-10861){\vector(-1, 0){2400}}
\put(1801,-9961){\vector( 1, 0){1500}}
\put(5701,-10411){\line( 1, 1){1800}}
\put(5701,-10411){\makebox(6.6667,10.0000){\SetFigFont{10}{12}{rm}.}}
\put(5701,-10411){\line( 1,-1){1800}}
\put(1801,-10861){\line( 1, 0){3300}}
\put(5101,-10861){\line( 1,-1){1800}}
\put(1801,-9961){\line( 1, 0){3300}}
\put(5101,-9961){\line( 1, 1){1800}}
\put(1801,-10861){\line( 1, 0){3300}}
\put(5101,-10861){\line( 1,-1){1800}}
\put(5101,-10861){\vector(-1, 0){2400}}
\put(1801,-9961){\line( 1, 0){3300}}
\put(5101,-9961){\line( 1, 1){1800}}
\put(5701,-10411){\vector( 1,-1){975}}
\put(7501,-8611){\vector(-1,-1){1050}}
\put(1801,-9961){\vector( 1, 0){1500}}
\put(5701,-10411){\line( 1, 1){1800}}
\put(5701,-10411){\makebox(6.6667,10.0000){\SetFigFont{10}{12}{rm}.}}
\put(5701,-10411){\line( 1,-1){1800}}
\put(-200,-10500){$\bk{\vecr}{nl}$}
\put(7500,-8500){$\bk{\vecr_1}{n_1l_1}$}
\put(7700,-12500){$\bk{\vecr_2}{n_2l_2}$}
\end{picture}

\caption{Strong decay topology.}
\label{topology}
\end{figure}
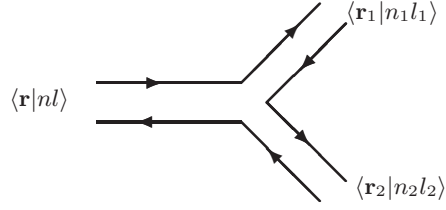

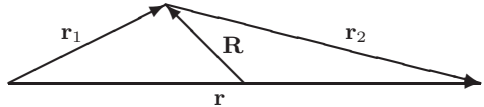
\begin{figure}[b]
%triangle.tex
\setlength{\unitlength}{0.000233300in}%
\begingroup\makeatletter\ifx\SetFigFont\undefined
% extract first six characters in \fmtname
\def\x#1#2#3#4#5#6#7\relax{\def\x{#1#2#3#4#5#6}}%
\expandafter\x\fmtname xxxxxx\relax \def\y{splain}%
\ifx\x\y   % LaTeX or SliTeX?
\gdef\SetFigFont#1#2#3{%
  \ifnum #1<17\tiny\else \ifnum #1<20\small\else
  \ifnum #1<24\normalsize\else \ifnum #1<29\large\else
  \ifnum #1<34\Large\else \ifnum #1<41\LARGE\else
     \huge\fi\fi\fi\fi\fi\fi
  \csname #3\endcsname}%
\else
\gdef\SetFigFont#1#2#3{\begingroup
  \count@#1\relax \ifnum 25<\count@\count@25\fi
  \def\x{\endgroup\@setsize\SetFigFont{#2pt}}%
  \expandafter\x
    \csname \romannumeral\the\count@ pt\expandafter\endcsname
    \csname @\romannumeral\the\count@ pt\endcsname
  \csname #3\endcsname}%
\fi
\fi\endgroup
\begin{picture}(10888,1888)(257,-10605)
\thicklines
\put(5001,-11061){$\vecr$}
\put(1500,- 9561){$\vecr_1$}
\put(5201,- 9861){$\vecR$}
\put(8001,- 9561){$\vecr_2$}
\put(301,-10561){\vector( 2, 1){3600}}
\put(3901,-8761){\vector( 4,-1){7200}}
\put(5701,-10561){\vector(-1, 1){1800}}
\put(301,-10561){\vector( 1, 0){10800}}
\end{picture}
\caption{Strong decay geometry.}
\label{geometry}
\end{figure}

In quark pair-creation processes  the dominant topology is that of FIG. \ref{topology}. Hairpin diagrams are generally suppressed or forbidden on account of colour considerations, or in the flux tube model because the created pair sit at the ends of two broken pieces of string constituting the final state hadrons. The corresponding geometry is show in FIG. \ref{geometry} where $\vecr,\vecr_1,\vecr_2$ are the $\qq$ coordinates of the initial and final states respectively, and 
\be
\vecR=\frac{\vecr_1-\vecr_2}{ 2},\qquad \vecr=\vecr_1+\vecr_2.
\label{likejacobi}
\ee
The spatial amplitude for such processes has integrals over the wavefunctions of the initial and final states with coordinates $\vecr, \vecr_1,\vecr_2$ and the relative wavefunction of the outgoing mesons in the relative coordinate $\vecr/2$. More generally, allowing for a transverse degree of freedom $\vecR$ (such as in a hybrid meson) the integrals are of the form
\be
\int \ud^3\vecr_1 \ud^3\vecr_2 \ud^3\vecr\ud^3\vecR \delta^3(\vecr-\vecr_1-\vecr_2) \delta^3(\vecR-\frac{\vecr_1-\vecr_2}{2})\ldots
\ee
Only two of the integrations need to be performed manually on account of the delta function: the only complication that arises is that at least one of the wavefunctions has to be expressed in the coordinates of another. This can be done using bipolar expansions but the presence of operators connecting the initial and final states, particularly in cases involving radially excited wavefunctions, makes the algebra unwieldy and masks underlying symmetries. A more convenient approach is to note that in its nature the problem is essentially the same as the earlier nuclear physics one: the wavefunctions must be translated from one set of Jacobi coordinates to another. Apart from a scale factor of $\sqrt 2$ the pair of coordinates \rf{likejacobi} is that obtained from the orthogonal coordinate transformation on the vectors $(\vecr_1,\vecr_2)$ cf. \rf{jacobi} above. Using the operator formalism of the harmonic oscillator  the matrix elements for such processes can be evaluated without having to perform any integration manually apart from simple radial integrals at the final stage.

The  approach is to first operate on the wavefunctions of the initial or final state, leaving wavefunctions in the coordinates $\vecr_1,\vecr_2$ and $\vecr$ (and in more general cases $\vecR$); those in the former coordinates are translated via the Talmi-Moshinsky transformation into a new set of wavefunctions of the latter coordinates,
\be
\bk{(n_1l_1\ot n_2l_2)_{l_{12}m_{12}}}{\vecr_1,\vecr_2}=
\sum_{NL nl}
\bk{(n_1l_1\ot n_2l_2)_{l_{12}      }}{(NL \ot nl)_{l_{12}      }}
\bk{(NL\ot nl)_{l_{12}m_{12}}}{\sqrt 2 \vecR,\vecr/{\sqrt 2}},
\label{talmi}
\ee
leaving only simple radial integrals. There are a finite number of ways of allocating the degrees of freedom to the transformed state, owing to the energy conservation rule \rf{energyconservation} and its corollary,
\be
(-)^{L+l}=(-)^{l_1+l_2}.
\label{orbconservation}
\ee
This approach is well suited to the problem of conventional and hybrid meson decay: the translated wavefunction has a unit of orbital angular momentum $L$ in the transverse degree of freedom $\vecR$, which must correspond to the initial transverse orbital angular momentum -- zero for a conventional meson or one for the first excited hybrid state. Along with the symmetries of the Talmi coefficients,
\be
\bk{(n_1l_1\ot n_2l_2)_{l_{12}      }}{(NL \ot nl)_{l_{12}      }}
=(-)^{l+l_{12}}\bk{(n_2l_2\ot n_1l_1)_{l_{12}      }}{(NL \ot nl)_{l_{12}      }},
\label{talmisymmetry}
\ee
this uncovers a new class of selection rules that probe the transverse degrees of freedom of the initial state, of which the  simplest example is the suppression of hybrid decays to pairs of identical S-wave mesons.

Most decay models are nonrelativistic in character and assume L, S factorise: the pictures differ in the assumed quantum numbers of the created pair. An early model assigned the  emergent pair the quantum numbers of the vaccum, corresponding to $\qq$ pair creation in a $\tpn$ state \cite{orsay}. With the arrival of QCD it was proposed that instead the $\qq$ should have the $\tso$ quantum numbers of the vector gluon but this was found to be less successful in describing the data \cite{geigerswanson}. In the strong coupling expansion of lattice QCD the natural degrees of freedom are not quarks and gluons but quarks and flux tubes and it is not a perturbative gluon that triggers decay but the breaking of a gluonic flux tube \cite{ip}. The Hamitonian allows for the quarks at the end of the broken pieces of string to carry either $\tpn$ or $\tso$ quantum numbers with corresponding operators:
\bea
&\tpn:\qquad &\mathbf{\chi}_1\cdot \grad =\mathbf{\chi}_1\cdot (\grad_{\vecr_1}+\grad_{\vecr_2}+\grad_{\vecr/2})\\
&\tso:\qquad &\mathbf{\chi}_1\cdot \hatr
\eea
where $\chi_1$ denotes the spin 1 wavefunction of the emergent pair and the vectors $\vecr$, $\vecr_1$ and $\vecr_2$ are those of FIG. \ref{geometry}. In the flux tube model of Isgur and Paton it is argued that the zero point oscillations of the string wash out the $\tso$ operator and that the $\tpn$ operator survives, converging on the phenomenologically successful $\tpn$ model and embedding it in a more appealing physical picture \cite{ki}.  An additional feature of the flux tube model is that the amplitude for decay is driven not only by the overlap of the quark wavefunctions but also by those of the flux tubes of each hadron. In the next section the amplitude for the $\tpn$ decay of any conventional meson into any other pair of conventional mesons will be formulated, neglecting the string degrees of freedom. For calculations involving hybrids the string degrees of freedom must be included: this is presented in the subsequent sections. 

\section{Conventional meson decay}
\label{sss}
The relevant degrees of freedom are  the colour, flavour, spin, orbital, radial and total angular momentum quantum numbers of each meson. The colour factor is the same for all processes of the type in  FIG. \ref{topology} and contributes only an overall factor which is absorbed into the normalisation. Flavour overlaps can be handled by standard techniques and will not be discussed here. For the initial meson with its $\qq$ axis at $\vecr$, the wavefunction assumes the form
\be
\bk{\vecr}{(s\ot nl)_{jm}}
\label{initial meson}
\ee
where $s$ corresponds to a pair of quarks whose labels have been suppressed coupled to spin $s$ and the space part is given by
\be
\bk{\mathbf{r}}{nlm_l}=\bk{r}{nl}Y_l^{m_l}(\mathbf{\hat r}).
\label{conventionalwavefunction}
\ee
The radial wavefunctions are taken as harmonic oscillators governed by a width parameter $\beta$, which in the present work is taken to be universal; typically $\beta\sim 0.4$GeV for light quarks. Corrections for wavefunctions of unequal width can be implemented at a later stage by mapping more realistic wavefunctions onto linear combinations of these harmonic oscillator basis states. Taking a universal wavefunction width suggests  a system of natural units in which $\beta=1$, so that the normalised radial wavefunction is 
\be
 \bk{r}{nl}=\mathcal{N}_{nl}r^l\mathcal{L}_n^{l+1/2}(r^2)e^{-r^2/2};\qquad \mathcal{N}_{nl}=\sqrt{\frac{2n!}{\Gamma(n+l+3/2)}}.
\ee
With the quantum numbers as per FIG. \ref{topology} and in the rest frame of the initial state, the final state consists of a pair of mesons $\bk{(s_1\ot n_1 l_1)_{j_1}}{\vecr_1}$ and  $\bk{(s_2\ot n_2 l_2)_{j_2}}{\vecr_2}$ with relative coordinate $\vecr/2$ moving apart with (dimensionless) momenta of magnitude $p$ measured in units of the wavefunction width $\beta$. The final states are coupled to $j_{12}$  and are in a relative partial wave $L$,
\be
\bk{LM_L}{\mathbf{r}/2}=\bk{r/2}{L}Y_L^{M_L}(\mathbf{\hat r}),
\label{relativemesonwavefunction}
\ee
where in the absence of final state interactions the relative radial wavefunction has the form
\be
\bk{L}{r/2}=\sqrt{\frac{2}{\pi}}i^Lj_L(pr/2).
\ee
The final state must be coupled to angular momentum $\ket{jm}$ on account of its being connected to the initial state by a scalar operator, so has the form 
\be
\bk{(((s_1\ot n_1 l_1)_{j_1}\ot(s_2\ot n_2 l_2)_{j_2})_{j_{12}}\ot L)_{jm}}{\vecr_1,\vecr_2,\vecr/2}
\ee
 The total width for the decay of a conventional meson into another pair of conventional mesons is the sum over couplings $j_{12}$ and relative partial waves $L$ of the squared amplitudes:
\be
\Gamma(snlj\to s_1n_1l_1j_1+s_2n_2l_2j_2)=2\pi\gamma_0^2\frac{pM_1M_2}{M}\sum_{j_{12}L}
\rme{(((s_1\ot n_1 l_1)_{j_1}\ot(s_2\ot n_2 l_2)_{j_2})_{j_{12}}\ot L)_j}{\mathbf{\chi}_1\cdot\grad}{(s\ot nl)_j}^2
\label{width}
\ee
where here $\gamma_0$ is a phenomenological pair creation constant. Details of choice of phase space will not be discussed in this paper, which is concerned only with the evaluation of the matrix element itself. In the matrix element above, and in subsequent manipulations, the state vectors $\ket{\vecr_1,\vecr_2,\vecr/2}$ and $\bra{\vecr}$ have been dropped and an overall integration
\be
\int \ud^3\vecr_1 \ud^3\vecr_2 \ud^3\vecr\ud^3\vecR \delta^3(\vecr-\vecr_1-\vecr_2) \delta^3(\vecR-\frac{\vecr_1-\vecr_2}{2})\ldots
\ee
is implied. The first step is to separate the spin and space degrees of freedom; the relevant recoupling is essentially the same as that of \cite{silvestrebrac} but it is useful to manipulate the radial quantum numbers along with the orbital parts,
\begin{multline}
\rme{(((s_1\ot n_1 l_1)_{j_1}\ot(s_2\ot n_2 l_2)_{j_2})_{j_{12}}\ot L)_j}{\mathbf{\chi}_1\cdot\grad}{(s\ot nl)_j}=
\sum_{ s_{12}l_{12}  l_f} 	(-)^{s+L+s_{12}+l_{12}+l_f}\st{l_fs_{12}l_{12}j_1j_2j_{12}}
\\
\ninej{s_1}{l_1}{j_1}{s_2}{l_2}{j_2}{s_{12}}{l_{12}}{j_{12}}
\sixj{s_{12}}{l_{12}}{j_{12}}{L}{j}{l_f}
\sixj{s_{12}}{s}{1}{l}{l_f}{j}
\rme{(s_1\ot s_2)_{s_{12}}}{\mathbf{\chi}_1}{s}
\rme{((n_1 l_1\ot n_2 l_2)_{l_{12}}\ot L)_{l_f}}{\grad}{nl}
\label{separatingspinandspace}
\end{multline}

with

\be
\st{ab\ldots}=\sqrt{(2a+1)(2b+1)\ldots}
\ee

The spin part is a 9-$j$ coefficient,
\be
\rme{(s_1\ot s_2)_{s_{12}}}{\mathbf{\chi}_1}{s}=\st{1ss_1s_2s_{12}}\ninej{\frac{1}{2}}{\frac{1}{2}}{s_1}{\frac{1}{2}}{\frac{1}{2}}{s_2}{1}{s}{s_{12}}.\label{spinpart}
\ee
The space part can be evaluated as follows:
\bea
\rme{((n_1 l_1\ot n_2 l_2)_{l_{12}}\ot L)_{l_f}}{\grad}{nl}&=&
\sum_{n_1'l_1'n_2'l_2'l_{12}'L'}
\rme{((n_1 l_1\ot n_2 l_2)_{l_{12}}\ot L)_{l_f}}
{\grad}
{((n_1'l_1'\ot n_2' l_2')_{l_{12}'}\ot L')_l}
\frac{1}{\st{l}}
\label{mmm1}
\\&&\qquad\qquad\times
\sum_{Nn'}
\bk{(n_1'l_1'\ot n_2'l_2')_{l_{12}'}}{(N0\ot n'l_{12}')_{l_{12}'}}
\label{mmm2}
\\&&\qquad\qquad\qquad\qquad\times
\bra{((N0\ot n'l_{12}')_{l_{12}'}\ot L' )_l}\ket{nl}.
\label{mmm3}
\eea
In the first step \rf{mmm1}, the operator $\grad$ acts on the final state wavefunctions to produce a new set of states which, to match the initial state ket, must be coupled to $l$. The operator $\grad$ is the sum of three terms $\grad_{\vecr_1},\grad_{\vecr_2},\grad_{\vecr/2}$ which act on the $\bra{n_1l_1}$, $\bra{n_2l_2}$ and $\bra{L}$ parts of the wavefunction respectively. The matrix element in \rf{mmm1} is found by recoupling to sandwich the operators between the relevant parts of the wavefunction:
\begin{multline}
\rme{((n_1l_1\ot n_2l_2)_{l_{12}}\ot L)_{l_f}}{\grad}{((n_1'l_1'\ot n_2' l_2')_{l_{12}'}\ot L')_l}=
\delta_{n_1l_1n_2l_2l_{12}}^{n_1'l_1'n_2'l_2'l_{12}'}
\st{l_f}(-)^{l_{12}+L'+l_f+1}\sixj{l_f}{1}{l}{L'}{l_{12}}{L}\rme{L}{\grad}{L'}
\\+(-)^{l_{12}+L+l}\st{l_fl_{12}l_{12}'}\sixj{l_f}{1}{l}{l_{12}'}{L}{l_{12}}
\Big(\delta_{      n_2l_2     L}^{        n_2'l_2'     L'}(-)^{l_1+l_2+l_{12}'}\sixj{l_{12}}{1}{l_{12}'}{l_1'}{l_2}{l_1}\rme{n_1l_1}{\grad}{n_1'l_1'}
\\+\delta_{      n_1l_1     L}^{        n_1'l_1'     L'}
(-)^{l_1+l_2'+l_{12}}\sixj{l_{12}}{1}{l_{12}'}{l_2'}{l_1}{l_2}\rme{n_2l_2}{\grad}{n_2'l_2'}
\Big).
\end{multline}
The matrix element $\rme{L}{\grad}{L'}$ can be found by operating on  $\bra L$ with the definition \rf{relativemesonwavefunction}: 
\bea
\rme{L}{\grad}{L-1}&=&ip\sqrt{L},\\
\rme{L}{\grad}{L+1}&=&-ip\sqrt{L+1}.
\eea
The matrix elements $\rme{n_1l_1}{\grad}{n_1'l_1'}$ and $\rme{n_2l_2}{\grad}{n_2'l_2'}$ follow from those of the  creation and annihilation operators,
\be
\veca=\frac{1}{\sqrt 2}(\vecp-i\vecr),\qquad \veca\dag=\frac{1}{\sqrt 2}(\vecp+i\vecr),
\label{creationannihilation}
\ee
so that  $\grad= i\vecp$ raises or lowers the number of quanta by one \cite{gs,haskellwybourne}. The matrix elements of $\vecr$ are also recorded here for later use,
\bea
\rme{nl}{\grad}{nl+1}=&\rme{nl}{\vecr}{nl+1}&=-\sqrt{(l+1)(n+l+3/2)},\label{me1}\\
\rme{nl}{\grad}{n+1l-1}=&\rme{nl}{\vecr}{n+1l-1}&=-\sqrt{l(n+1)},\label{me2}
\eea
and 
\bea
\rme{n'l'}{\grad}{nl}&=&\rme{nl}{\grad}{n'l'}\label{me1flip}\\
\rme{n'l'}{\vecr}{nl}&=&-\rme{nl}{\mathbf{\vecr}}{n'l'}\label{me2flip}.
\eea
In the second step \rf{mmm2}, the harmonic oscillator states in $\ket{\vecr_1,\vecr_2}$ are translated into states in $\ket{\sqrt 2\vecR,\vecr/\sqrt 2}$, as in EQN. \rf{talmi}, 
\be
\bk{(n_1'l_1'\ot n_2'l_2')_{l_{12}'m_{12}'}}{\vecr_1,\vecr_2}=
\sum_{N n'}
\bk{(n_1'l_1'\ot n_2'l_2')_{l_{12}' }}{(N0 \ot n'l_{12}')_{l_{12}'      }}
\bk{(N0\ot n'l_{12}')_{l_{12}'m_{12}'}}{\sqrt 2 \vecR,\vecr/{\sqrt 2}}.
\ee
Here the orbital angular momentum in the harmonic oscillator state in $\sqrt 2\vecR$ is forced to be zero on account of the initial state $\ket {nl}$ having zero orbital angular momentum in that direction; this leaves orbital angular momentum $l_{12}'$ in the direction $\vecr$. What remains is a sum over radial quantum numbers $n',N$ restricted by the energy conservation rule \rf{energyconservation}.
This takes care of the integral
\be
\int \ud^3\vecr_1 \ud^3\vecr_2  \delta^3(\vecr-\vecr_1-\vecr_2) \delta^3(\vecR-\frac{\vecr_1-\vecr_2}{2})
\ee
leaving only an integral  over $\ud^3\vecr\ud^3\vecR$, represented by the final line \rf{mmm3}; the integral over $\ud\hatR$ is trivially one, so that
\be
\bra{((N0\ot n'l_{12}')_{l_{12}'}\ot L')_l}\ket{nl}
=
\bra{l_{12}' L'}\ket{l}
\int_0^\infty \ud RR^2\bk{N0}{\sqrt{2}R}
\int_0^\infty \ud rr^2\bk{n'l_{12}' }{r/\sqrt{2}}\bk{L'}{r/2}\bk{r}{nl}
\label{manualintegrations}
\ee
with
\bea
\bra{l_{12}' L' }\ket{l}&=&\st{l_{12}' L' }\cleb{l_{12}' }{0}{L' }{0}{l}{0}\label{orbitalr}\\
\int_0^\infty \ud RR^2\bk{N0}{\sqrt{2}R}&=&\frac{\sqrt \pi}{4}\sum_{K=0}^{N}a_{N0K}(1+2K)!!
\label{radialR}\\
\int_0^\infty \ud rr^2\bk{n'l_{12}' }{r/\sqrt{2}}\bk{r/2}{L'}\bk{r}{nl}
&=&\frac{i^{L' }}{\sqrt 2}\left(\frac{p}{4}\right)^{L'}
\sum_{kk'=0}^{nn'}\frac{a_{nlk}a_{n'l_{12}'k'}}{2^{k'+l_{12}' /2}}\left(\frac{4}{3}\right)^{\frac{\alpha}{2}}
\frac{\Gamma(\frac{\alpha}{2})}{\Gamma(L' +\frac{3}{2})}
\mathcal{M}\left(\frac{\alpha}{2},L'+\frac{3}{2},-\frac{p^2}{12}\right)
\label{radialr}
\eea
where the index $\alpha=L' +l+2k+l_{12}' +2k'+3$ and coefficients from the expansion of the Laguerre polynomials are
\be
a_{nlk}=\mathcal{N}_{nl}\frac{(-)^k}{k!}
\frac{\Gamma(n+l+3/2)}{(n-k)!\Gamma(k+l+3/2)}.
\ee 
The resulting expressions have a centrifugal barrier suppressing higher partial waves at small momentum  and a form factor,
\be
p^L\times(\textrm{polynomial in } p^2)\times e^{-\frac{p^2}{12}}.
\label{polynomialform}
\ee
The polynomial form can be generated by repeated application of the confluent hypergeometric function recursion relation
\be
\mathcal M(a,b,-\frac{p^2}{12})=\mathcal M(a-1,b,-\frac{p^2}{12})-\frac{p^2}{12b}\mathcal M(a,b+1,-\frac{p^2}{12})
\ee 
and using $\mathcal M (a,a,-\frac{p^2}{12})=e^{-\frac{p^2}{12}}$; this is always possible because the first two indices in $\mathcal M$ will differ by an integer as \rf{orbitalr} demands  $l_{12}'+L'+l$ is even. The expressions so obtained recover the results of REFS. \cite{ki} and \cite{bcps} and extend upon them to consider decays not treated there, including modes that are not relevant the decays of physical mesons but are useful in modelling  virtual $\qq$ decays \cite{ccdecay}. 

%%%%%%%%%%%%%%%%%%%%%%%%%%%%%%%%%%%%%%%%%%%%%%%%%%%%%%%%%%%%%%%%%%%%%%%%%%%%%%%%%%%%%%%%%%%%%%%%%%%%%%%%%%%%%%%

\section{Mesons with flux tubes}
\label{effwavefns}

In the flux tube model the state of the gluon field is manifest: a meson with angular momentum $l$ is built from the quark ($l'$) and flux tube ($l_s$) angular momentum
\be
\vecl=\vecl'+\vecls.\label{lquarkpluslstring}
\ee
Because of the axial symmetry of the flux tube part of the wavefunction,  the full wavefunction is not in general an eigenstate  of $\vecl'^2$ or $\vecls^2$ and there are no associated quantum numbers $l',l_s$. On the other hand, the projection of the angular momentum along the molecular  axis is conserved quantity: for a state with $\qq$ axis  at $\hatr$ the relevant operator is $\vecl\cdot\hatr=\vecl_s\cdot\hatr$ and the associated eigenvalue is  $\lambar$. The states with nonzero $\lambar$ are gluonic hybrids and the $\pm\lambar$ states are degenerate left- and right-moving phonons, the interference of which gives parity doublets. The spatial wavefunctions can be described by kets of the form
\be
\ket{n\lam l^Pm_l}
\ee
where here $n$ is the quark radial quantum number, $\lam=|\lambar|$, $l$ is the total orbital angular momentum restricted to $l\ge\lam$, and $P$ is a parity label relevant only to the $\lam\ne 0$ states. In the parlance of molecular physics, states with $\lam=0,1, 2\ldots$ are labeled $\Sigma,\Pi,\Delta\ldots$ in analogy with S,P,D\ldots, the latter being used to label the quantum number $l$. The $\Sigma$ states are the conventional quark model states; combining with spin gives the $J^{PC}$ quantum numbers:
\bea
\ket{n^1\Sigma\uS_0}	\qquad0^{-+}	&\qquad\ket{n^3\Sigma\uS_1}		&\qquad1^{--}	\\
\ket{n^1\Sigma\uP_1}	\qquad1^{+-}	&\qquad\ket{n^3\Sigma\uP_{0,1,2}}	&\qquad0,1,2^{++}\\
\ket{n^1\Sigma\uD_2}	\qquad2^{-+}	&\qquad\ket{n^3\Sigma\uD_{1,2,3}}	&\qquad1,2,3^{--}\qquad\&c.
\eea
The lowest lying orbital configurations for hybrid states are the $\Pi\uP^\pm$ states, with a single phonon in the lowest mode of excitation of the flux tube and total orbital angular momentum $l=1$. Combining with spin this set includes the manifestly exotic $0^{+-},1^{-+},2^{+-}$ states:
\bea
\ket{n^1\Pi\uP^+_1}	\qquad1^{++}	&\qquad&\ket{n^3\Pi \uP^+_{0,1,2}}	\qquad0,1,2^{+-}\\
\ket{n^1\Pi\uP^-_1}	\qquad1^{--}	&\qquad&\ket{n^3\Pi \uP^-_{0,1,2}}	\qquad0,1,2^{-+}
\eea
Those of most phenomenological interest are the quark radial ground states denoted $\ket{1\Pi\uP^\pm}$ where here the prefix ``1'' stands for $n+1$ as in the usual chemical notation. These are the family of hybrids which for light quarks are expected to lie at around 1.8--2.0GeV and are discussed in REFS. \cite{ikp,cp95}. As well as radial excitations $\ket{2\Pi\uP^\pm}$\&c., there exist in principle orbital excitations such as those with $l=2$,
\bea
\ket{n^1\Pi\uD^+_2}	\qquad2^{++}	&\qquad&\ket{n^3\Pi\uD^+_{1,2,3}}	\qquad1,2,3^{+-}\\
\ket{n^1\Pi\uD^+_2}	\qquad2^{--}	&\qquad&\ket{n^3\Pi \uD^-_{1,2,3}}	\qquad1,2,3^{-+}
\eea
and then again a sequence of states with two phonons in the flux-tube $\ket{n\Delta\uD^\pm},\ket{n\Delta\uF^\pm}$\&c.\footnote{Degenerate with these  $\Delta$-type hybrids are a set of $\Sigma'$ states with two phonons coupled to $\lam=0$; these will not be discussed further here.}
%All of the $\Pi$ states with non-exotic $J^{PC}$ have opposite spin to the $\Sigma$ states which share their quantum numbers. 

With the flux-tube degrees of freedom manifest  the amplitude for a decay depends not only on the overlap of the quark wavefunctions but also on those of the flux tubes of the initial and final hadrons. Dowrick, Paton and Perantonis\cite{dpp} obtain the flux tube overlap  for decays of the type
\bea
\Sigma&	\to&\Sigma+\Sigma\textrm{, and}\\
\Pi   &	\to&\Sigma+\Sigma	\label{pss}
\eea
for a  harmonic flux tube connecting a fixed quark and antiquark. These factors provide a parameter-free scale for processes involving hybrids relative to those involving only conventional mesons, driven by calculable terms $\kappa$ and the ratio $\sqrt \sigma/\beta$ where $\sigma$ is the string tension. For a wide range of other decay processes, including 
\bea
\Sigma&	\to&\Pi+\Sigma,		\\	
\Sigma&	\to&\Pi+\Pi\textrm{, and}		\\	
\Delta&	\to&\Sigma+\Sigma	\label{dss}
\eea
the string overlap factors share the same generic form\footnote{The flux tube overlap factors relevant to decays with hybrids in both the initial and final states contain additional terms to those described and will not be discussed in any detail here. }: string breaking is suppressed exponentially according to the perpendicular distance $\vecrho$ away from the $\qq$ axis, localising  pair creation to a region of order $1/\sqrt \sigma$, and there are solid harmonic factors in $\vecrho$ associated with the gluonic excitations $\Pi,\Delta,\&c.$   

In the adiabatic approach  the string overlap factors are used as a spatial weight for the full decay amplitude \cite{ki,ikp}. The subsequent calculations in this paper, which follow the prescription of REF. \cite{tjbunpublished}, work in the approximation that the exponential localisation of string breaking is discarded. In this limit decay amplitudes involving only $\Sigma$ states  are those of the naive  $\tpn$ model in which the amplitude for pair creation is equal everywhere in space; this is known to be a good approximation, essentially  because the quark wavefunctions mimick this asymptotic form \cite{ki}. This is the level of approximation within which the amplitudes presented in \cite{ftlattice} were calculated, and the agreement with the lattice was found to be good. 

What remains of each string overlap is a solid harmonic associated with each gluonic excitation: in this sense the string overlaps are factorisable. In the approach of REF.  \cite{tjbunpublished} the appropriate factor is combined with the quark wavefunction of REF. \cite{ip} to give an effective quark-flux tube wavefunction for each state; the overall decay amplitude is then driven by the factors $\kappa$ which are also calculated there. It is found that the combined quark-flux tube wavefunction can be expressed as  a linear combination of tensors of rank $l$,
\be
\bk{\vecrho,\vecr}{(\lam\ot n'l')_{lm_l}}=\sum_{\mu m'}\bk{\vecrho}{\lam \mu}\bk{\mathbf{\vecr}}{n'l'm'}\bk{\lam \mu,l'm'}{lm_l}.
\label{effwavefn}
\ee
where the $\qq$ degree of freedom is a harmonic oscillator as before
\be
\bk{\mathbf{r}}{n'l'm'}=\bk{r}{n'l'}Y_{l'}^{m'}(\mathbf{\hat r})
\label{conventionalwavefunctionagain}
\ee
and the flux-tube part of the wavefunction is 
\be
\bk{\vecrho}{\lam \mu}=\sqrt{\frac{4\pi}{2\lam+1}}\rho^\lam Y_{\lam}^{\mu}(\hatrho)\label{ftsolidharmonic}.
\ee
The translation between the full wavefunction and the effective wavefunction,
\be
\ket{n\lam l^P m_l}\to \sum_{n'l'}\ket{(\lam\ot n'l')_{lm_l}}\bk{(\lam\ot n'l')_l}{n\lam l^P},
\ee
is discussed in detail in REF. \cite{tjbunpublished}: here only the final results are quoted for the cases of most interest. For conventional mesons ($\lam=\Sigma$) the translation is trivial,
\bea
\ket{n\Sigma lm_l}&\to&\ket{ (\Sigma\ot n l)_{lm_l} }\\
	       &=&\ket{nlm_l}
\eea
where in the second step, which follows from  the definition \rf{effwavefn}, there appears a harmonic oscillator ket defined in EQN. \rf{conventionalwavefunction}. Amplitudes involving only conventional mesons are then the amplitudes of the old $\tpn$ model with no string degrees of freedom, corresponding to pair creation with equal amplitude everywhere in space. 
%\be
%\bk{\vecrho,\vecr}{(\Sigma\ot n'l')_{lm}}=\bk{\vecr}{n'l'm'}\delta_{ll'}\delta_{mm'}
%\ee

For hybrid states ($\lam=\Pi,\Delta,\ldots$) the effective wavefunctions are, unlike the full quark-flux tube wavefunctions, eigenstates of both $\vecl'^2$ and $\vecls^2$:
\bea
\vecl'^2\ket{(\lam\ot n'l')_{lm_l}}&=&l'(l'+1)\ket{(\lam\ot n'l')_{lm_l}},\\
\vecl_s^2\ket{(\lam\ot n'l')_{lm_l}}&=&\lam(\lam+1)\ket{(\lam\ot n'l')_{lm_l}}.
\eea
This is a consequence of having discarded the axially symmetric exponential localisation of string breaking. Thus $\Sigma,\Pi,\Delta\ldots$ states have string angular momentum equal to the magnitude of its projection along the molecular axis, $l_s=\lam=0,1,2,\ldots$ The quark orbital angular momentum $l'$ can take values $|\lam-l|\le l'\le \lam+l$ and the parity of a state is  $(-)^{\lam+l'+1}$, so that the parity doublets correspond to $l'$ even/odd. The quark radial wavefunctions for hybrids are subject to a modified centrifugal barrier, but decay calculations appear to be fairly insensitive to this \cite{cp95}. In the approximation that the radial wavefunction is a 1P harmonic oscillator state, the effective wavefunctions for the lowest lying hybrids are found to be
% States can be built with any $l>\lam$ and any quark radial excitation, the lightest being the $\Pi$ state coupled to $l=1$ and with quark radial ground state. These are the familiar hybrid states expected to lie around 1.8--2.0GeV and include the manifestly exotic $0^{+-},1^{-+},2^{+-}$ states. 
\bea
\ket{1\Pi  \uP^-} 	&\to&\ket{(\Pi\ot 1\uP)_1}	\label{pim1P}\\
\ket{1\Pi  \uP^+}	&\to&\sqrt{\frac{2}{3}}\left(2\sqrt{\frac{2}{3\pi}}\ket{(\Pi\ot 1\uS)_1}+\ldots\right)
+\sqrt{\frac{1}{3}}\left(\frac{8}{3}\sqrt{\frac{2}{5\pi}}\ket{(\Pi\ot 1\uD)_1}+\ldots\right)		\label{pip1P}
\eea
The ellipsis indicate contributions from higher radial excitations in the quantum number $n'$, owing to the mismatch between the orbital and radial characters of the quark part of the wavefunctions. For the orbitally excited states it is those with negative parity that pick up correction terms for the radial wavefunctions, which is assumed to be 1D:
\bea
\ket{1\Pi  \uD^-} &\to&	 \sqrt{\frac{3}{5}}\left(
						\frac{8}{3}\sqrt{\frac{2}{5\pi}}\ket{(\Pi\ot 1\uP)_2}+\ldots\right)
						+\sqrt{\frac{2}{5}}\left(\frac{16}{5}\sqrt{\frac{2}{7\pi}}\ket{(\Pi\ot 1\uF)_2}
						+\ldots\right) \label{pim1D}\\
\ket{1\Pi  \uD^+} &\to& \ket{(\Pi\ot 1\uD)_2}\label{pip1D} 		
\eea
In a similar way it is possible to write down wavefunctions for $\Delta$-type hybrids and so forth.

\begin{figure}
%qpc.tex
\setlength{\unitlength}{0.00023300in}%
\begingroup\makeatletter\ifx\SetFigFont\undefined
% extract first six characters in \fmtname
\def\x#1#2#3#4#5#6#7\relax{\def\x{#1#2#3#4#5#6}}%
\expandafter\x\fmtname xxxxxx\relax \def\y{splain}%
\ifx\x\y   % LaTeX or SliTeX?
\gdef\SetFigFont#1#2#3{%
  \ifnum #1<17\tiny\else \ifnum #1<20\small\else
  \ifnum #1<24\normalsize\else \ifnum #1<29\large\else
  \ifnum #1<34\Large\else \ifnum #1<41\LARGE\else
     \huge\fi\fi\fi\fi\fi\fi
  \csname #3\endcsname}%
\else
\gdef\SetFigFont#1#2#3{\begingroup
  \count@#1\relax \ifnum 25<\count@\count@25\fi
  \def\x{\endgroup\@setsize\SetFigFont{#2pt}}%
  \expandafter\x
    \csname \romannumeral\the\count@ pt\expandafter\endcsname
    \csname @\romannumeral\the\count@ pt\endcsname
  \csname #3\endcsname}%
\fi
\fi\endgroup
\begin{picture}(5788,4588)(1757,-12705)
\thicklines
\put(5401,-9661){\vector( 1, 1){1050}}
\put(6676,-12436){\vector(-1, 1){975}}
\put(5701,-10411){\vector( 1,-1){975}}
\put(7501,-8611){\vector(-1,-1){1050}}
\put(5101,-10861){\vector(-1, 0){2400}}
\put(1801,-9961){\vector( 1, 0){1500}}
\put(5701,-10411){\line( 1, 1){1800}}
\put(5701,-10411){\makebox(6.6667,10.0000){\SetFigFont{10}{12}{rm}.}}
\put(5701,-10411){\line( 1,-1){1800}}
\put(1801,-10861){\line( 1, 0){3300}}
\put(5101,-10861){\line( 1,-1){1800}}
\put(1801,-9961){\line( 1, 0){3300}}
\put(5101,-9961){\line( 1, 1){1800}}
\put(1801,-10861){\line( 1, 0){3300}}
\put(5101,-10861){\line( 1,-1){1800}}
\put(5101,-10861){\vector(-1, 0){2400}}
\put(1801,-9961){\line( 1, 0){3300}}
\put(5101,-9961){\line( 1, 1){1800}}
\put(5701,-10411){\vector( 1,-1){975}}
\put(7501,-8611){\vector(-1,-1){1050}}
\put(1801,-9961){\vector( 1, 0){1500}}
\put(5701,-10411){\line( 1, 1){1800}}
\put(5701,-10411){\makebox(6.6667,10.0000){\SetFigFont{10}{12}{rm}.}}
\put(5701,-10411){\line( 1,-1){1800}}
\put(-2600,-10500){$\bk{\vecR,\vecr}{(\lam\ot n'l')_{lm}}$}
\put(7500,-8500){$\bk{\vecr_2,\vecr_1}{(\lam_1\ot n_1'l_1')_{l_1m_1}}$}
\put(7700,-12500){$\bk{\vecr_1,\vecr_2}{(\lam_2\ot n_2'l_2')_{l_2m_2}}$}
\end{picture}

\caption{State vectors and quantum numbers.}
\label{topology2}
\end{figure}
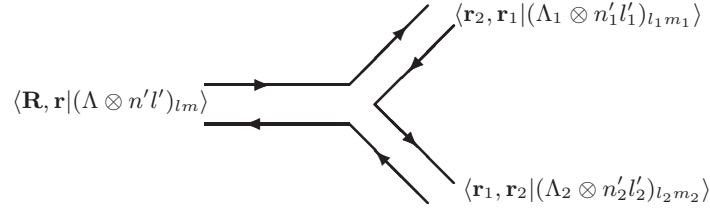

Working in this basis, the decay amplitude that appears in equation \rf{width} has the form
\be
\rme{(((s_1\ot (\lam_1\ot n_1'l_1')_{l_1})_{j_1}\ot(s_2\ot (\lam_2\ot n_2'l_2')_{l_2})_{j_2})_{j_{12}}\ot L)_j}
{\mathbf{\chi}_1\cdot\grad}
{(s\ot (\lam\ot n'l')_l)_j}.
\ee
The spatial parts of the meson wavefunctions are still tensors of rank $l,l_1,l_2$ so the recoupling in equation \rf{separatingspinandspace} to separate the spin and space degrees of freedom is the same and the task is to evaluate matrix elements of the form
\be
\rme{(((\lam_1\ot n_1'l_1')_{l_1}\ot(\lam_2\ot n_2'l_2')_{l_2})_{l_{12}}\ot L)_{l_f}}{\grad}{(\lam\ot n'l')_l}.\label{requisiteme}
\ee
In formulating the full decay amplitude it is useful to associate $\vecrho$ with one of the vectors appearing in FIG. \ref{geometry}. As discussed elsewhere \cite{tjbunpublished}, the purely transverse oscillations of the string allows the replacement
\be
\vecrho\to\vecR
\ee
for the flux tube vector of the initial state whose $\qq$ axis is at $\vecr$. Within the  harmonic approximation the same replacement can be made for  flux tube excitations in the final state, and because the flux tube has only transverse oscillations this is equivalent to
\bea
\vecrho&\to&\vecr_2/2,\\
\vecrho&\to&\vecr_1/2,
\eea
for the final state mesons whose $\qq$ axes are at $\vecr_1,\vecr_2$ respectively. The quantum numbers  and dynamical variables associated with each state are summarised in FIG. \ref{topology2}.
The expansion \rf{effwavefn} need never be done explicitly, and it is more useful to work in the coupled form $\bk{\vecrho,\vecr}{(\lam\ot n'l')_{lm}}$. The solid harmonic \rf{ftsolidharmonic} describes the elements of a tensor of rank $\lam$, so that for  
\be
\Sigma,\quad\Pi,\quad\Delta\ldots
\ee 
states the tensor $\lam$ corresponds to
\be
1,\quad\vecrho,\quad\{\vecrho\ot\vecrho\}_2,\ldots
\ee
Thus apart from a scale factor the tensors associated with each gluonic excitation are:
\begin{center}
\begin{tabular}{llccc}
\hline
	&	&$\lam$		&$\lam_1$		&$\lam_2$	\\	
\hline
$\Sigma$&\qquad	&1		&1		&1		\\
$\Pi   $&\qquad	&$\vecR$		&$\vecr_2$	&$\vecr_1$	\\
$\Delta$&\qquad	&$\{\vecR\ot\vecR\}_2$		
					&$\{\vecr_2\ot\vecr_2\}_2$	
								&$\{\vecr_1\ot\vecr_1\}_2$\\
\hline
\end{tabular}
\end{center}
%The convenience of this approach is that these tensors can be treated as operators on the harmonic oscillator degrees of freedom that already exist in the problem.  
For the final states $\lam_1,\lam_2$ correspond respectively to tensors in $\vecr_2,\vecr_1$ which can be treated  as operators on the $\ket{n_2'l_2'}$ and $\ket{n_1'l_1'}$ harmonic oscillator degrees of freedom that appear in \rf{requisiteme}; this case is discussed in the next but one section. For an initial state hybrid $\lam$ excitations are treated as operators on a harmonic oscillator degree of freedom in $\vecR$ that appears in the Talmi-Moshinsky transformation; this is discussed in the next section.  For $\Sigma$ states the matrix elements are just those of the unit operator,
\be
\rme{nl}{\Sigma}{n'l'}=\delta_{nn'}\delta_{ll'}\st{l}\label{ho1}.
\ee
For the  $\Pi$ states the requisite matrix elements are those of the vector $\vecrho$, already given in \rf{me1}--\rf{me2flip}: 
\bea
\rme{nl}{\Pi}{nl+1}&=&-\sqrt{(l+1)(n+l+3/2)},\\
\rme{nl}{\Pi}{n+1l-1}&=&-\sqrt{l(n+1)}\textrm{, and}\\
\rme{n'l'}{\Pi}{nl}&=&-\rme{nl}{\Pi}{n'l'}.\label{ho2}
\eea
For $\lam>1$ hybrids the reduced matrix elements of the  corresponding solid harmonic are required; these are known but such cases will not be discussed any further here.
%%%%%%%%%%%%%%%%%%%%%%%%%%%%%%%%%%%%%%%%%%%%%%%%%%%%%%%%%%%%%%%%%%%%%%%%%%%%%%%%%%%%%%%%%%%%%%%%
\section{Hybrid meson decay}
\label{lss}

In this section the spatial part of the amplitude for the decay of a state of arbitrary gluonic angular momentum into a pair of conventional mesons
\be
\lam\to\Sigma+\Sigma,
\ee
 will be formulated, recovering the results of the previous section if the initial meson has $\lam=\Sigma$. The  amplitude can be expressed in a form analagous to that of the conventional mesons case \rf{mmm1}--\rf{mmm3}:
\bea
\rme{(((n_1  l_1  \ot n_2  l_2  )_{l_{12}})\ot L)_{l_f}}
{\grad}
{(\lam\ot n'l')_l}&=&
\sum_{n_1'l_1'n_2'l_2'l_{12}'L'}
\rme{((n_1 l_1\ot n_2 l_2)_{l_{12}}\ot L)_{l_f}}
{\grad}
{((n_1'l_1'\ot n_2' l_2')_{l_{12}'}\ot L')_l}
\frac{1}{\st{l}}
\label{hmm1}
\\&&\qquad\qquad\times
\sum_{Nn''l''}
\bk{(n_1'l_1'\ot n_2'l_2')_{l_{12}'}}{(N\lam\ot n''l'')_{l_{12}'}}
\label{hmm2}
\\&&\qquad\qquad\qquad\qquad
\times\bra{((N\lam\ot n''l'')_{l_{12}'}\ot L)_l}
\ket{(\lam\ot n'l')_l}.
\label{hmm3}
\eea
Here the first step \rf{hmm1} is the same as for the conventional meson case \rf{mmm1}. In the second  step the harmonic oscillator states in $\ket{\vecr_1,\vecr_2}$ are translated into states in $\ket{\sqrt 2\vecR,\vecr/\sqrt 2}$, as before, but now the translated states must have angular momentum $\lam$ in the direction $\vecR$ to match the initial state $\ket{(\lam\ot n'l')_l}$:
\be
\bk{(n_1'l_1'\ot n_2'l_2')_{l_{12}'m_{12}'}}{\vecr_1,\vecr_2}=
\sum_{N n''l''}
\bk{(n_1'l_1'\ot n_2'l_2')_{l_{12}' }}{(N\lam \ot n''l''   )_{l_{12}'      }}
\bk{(N\lam\ot n''l''    )_{l_{12}'m_{12}'}}{\sqrt 2 \vecR,\vecr/{\sqrt 2}}.
\ee
The matrix element in the final step can be integrated directly,  but it is more convenient to treat $\lam$ as an operator on the $\vecR$ degree of freedom and make use of the  integrations already performed,
\be
\bra{((N\lam\ot n''l'')_{l_{12} }\ot L)_l}
\ket{(\lam\ot n'l')_l}
=
(-)^{L+l'+l''}\frac{\Pi_{ll_{12}}}{\Pi_{\lam}}\sixj{l}{\lam}{l'}{l''}{L}{l_{12}}
\sum_{N'}
\rme{N\lam}{\lam}{N'0}
\bra{((N'0\ot n''l'')_{l''}\ot L)_{l'}}\ket{n'l'}
\label{lastline}
\ee
%THE BELOW IS MORE COMPLETE BUT PROBABLY EXCESSIVELY SO,
%\bea
%\bra{((N\lam\ot n''l'')_{l_{12} }\ot L)_l}
%\ket{(\lam\ot n'l')_l}
%&=&
%\rme{((N\lam\ot n''l'')_{l_{12} }\ot L)_l}{\lam}{n'l'}\\
%&=&
%\sum_{N'}\rme{((N\lam\ot n''l'')_{l_{12} }\ot L)_l}{\lam}{((N'0\ot n''l'')_{l'' }\ot L)_{l'}}
%\bra{((N'0\ot n''l'')_{l'' }\ot L)_{l'}}{n'l'}\\
%&=&(-)^{L+l'+l''}\frac{\Pi_{ll_{12}}}{\Pi_{\lam}}\sixj{l}{\lam}{l'}{l''}{L}{l_{12}}
%\sum_{N'}
%\rme{N\lam}{\lam}{N'0}
%\bra{((N'0\ot n''l'')_{l''}\ot L)_{l'}}\ket{n'l'}
%\label{lastline}
%\eea
where the harmonic oscillator matrix element is given by EQNS. \rf{ho1}--\rf{ho2} and the final term is given by EQN. \rf{manualintegrations}. It is easily checked that if the initial state is a conventional meson then the expressions \rf{hmm1}--\rf{hmm3} collapse to the results of the previous section \rf{mmm1}--\rf{mmm3}.

The amplitudes so obtained encompass many modes not yet discussed in the literature: these are too numerous to record here in their final polynomial form and so will be collated and discussed elsewhere. These hybrid decay amplitudes have the same generic functional form \rf{polynomialform} as the conventional meson amplitudes. The decays
\be
1\Pi\uP^\pm\to 1\uP+^1\uS_0,
\ee
reproduce the results of REF. \cite{cp95}\footnote{The expression for $1^{-+}\to 1\uP+1\uS$ quoted in \cite{ftlattice} corrects a numerical error in \cite{cp95}.}, where here and in what follows the label $\Sigma$ for conventional mesons is omitted. The overall scale for these decays is set by the string overlap term; that scale factor is now being tested on the lattice and the agreement is good \cite{mm06,ftlattice}. The authors of REF. \cite{cp95} found closed forms for decays of the negative parity states $1\Pi\uP^-$  but not for those with  positive parity $1\Pi\uP^+$. This arises from the mismatch between the orbital and radial characters of the wavefunctions, explicit in the wavefunctions \rf{pim1P} and \rf{pip1P}:    the amplitudes derived here for $1\Pi\uP^-$ states are finite polynomials in $p$ whereas those for $1\Pi\uP^+$ states are a (rapidly decreasing) power series in $p$. 

The amplitudes derived here extend upon previous results in several different directions. The results are for arbitrary radial quantum number $n'$ and so can be applied to  more realistic radial wavefunctions that take account of the modified centrifugal barrier for hybrid states: any such wavefunction can be expressed as a linear combination of radial harmonic oscillator states. For the $1\Pi\uP^\pm$ states modes such as 
\bea
1\Pi \uP^\pm&\to&1\uP+\tso \textrm{, and}\\
1\Pi \uP^\pm&\to&1\uD+\ssn
\eea
 have not previously been considered and have implications for hybrid width if hybrid mass is sufficient to allow decay \cite{tjbunpublished}; it may also be possible to test these on the lattice where masses can be adjusted to allow decay. The amplitudes derived here also include more exotic classes of decays that may be testable on the lattice, such as the decay of orbitally excited $\Pi$ hybrids or $\Delta $ hybrids:
\bea
1\Pi \uD^\pm&\to&\Sigma+\Sigma,\\
1\Delta\uD^\pm&\to&\Sigma+\Sigma \textrm{ \&c., }
\eea
where here $\Sigma$ stands generically for any conventional meson. Returning to the lightest $1\Pi\uP^\pm$ states, the operator formalism admits a simple proof of the selection rule of Close and Page \cite{cp95} forbidding their decay to pairs of 1S mesons, 
\be
1\Pi\uP^\pm\nrightarrow 1\uS+1\uS\label{pi1S1S}.
\ee
Acting on a final state pair of identical 1S mesons  the decay operator is proportional to $\vecr$, using EQN. \rf{me1}:
\bea
\rme{((1\uS\ot 1\uS)_{  0}\ot L)_{L  }}{\grad}{(\lam\ot n'l')_l}
&=&\rme{((1\uS\ot 1\uS)_{  0}\ot L)_{L  }}{\grad_{\vecr_1}+\grad_{\vecr_2}+ \grad_{\vecr/2}}{(\lam \ot n'l')_l}\\
&=&\rme{((1\uS\ot 1\uS)_{  0}\ot L)_{L  }}{{\vecr_1}+{\vecr_2}+ \grad_{\vecr/2}}{(\lam \ot n'l')_l}\\
&=&\rme{((1\uS\ot 1\uS)_{  0}\ot L)_{L  }}{\vecr+ \grad_{\vecr/2}}{(\lam \ot n'l')_l}
\eea
Since neither the final state nor the operator have angular momentum about  $\vecR$, the initial state can only have zero angular momentum about $\vecR$, corresponding to a $\Sigma$ state: thus the decay of any $\lam\ne 0$ state to a pair of 1S mesons is forbidden:
\be
\rme{(1\uS\ot 1\uS)_{ 0}\ot L)_{L  }}{\grad}{(\lam\ot n'l')_l}=0 \textrm{ for } \lam\ne 0.
\ee
The analogue of \rf{pi1S1S}  can now be expressed in a  more general form which is independent of the radial wavefunction and the total angular momentum $l$ of the initial state:
\be
\rme{(1\uS\ot 1\uS)_{  0}\ot L)_{L  }}{\grad}{(\Pi\ot n'l')_l}=0,\label{rathermoregeneral}
\ee
Since the zero appears for any $n'$ it holds for any linear combination of $n'$ such as a more realistic wavefunction which takes account of the modified centrifugal barrier for gluonic excitations. The result is also independent of the total orbital angular momentum $l$, forbidding the decay of not just $\Pi\uP^\pm$ states but also their orbitally excited counterparts $\Pi \uD^\pm$ \&c. As noted in REFS. \cite{ikp,cp95}, the rule is broken by outgoing mesons with different spatial wavefunctions: this can be understood by returning to physical units with different wavefunction widths $\beta_1\ne\beta_2$, in which case the decay operator has a part which is proportion to $(\beta_1^2-\beta_2^2)\vecR$ and the decay from a $\Pi$ state is allowed. A new result is that the rule also forbids such decays for hybrids with two phonons,
\be
\rme{(1\uS\ot 1\uS)_{ 0}\ot L)_{L  }}{\grad}{(\Delta\ot n'l')_l}=0,
\ee
and so forth. %The authors of REF. \cite{ikp} offer a semiclassical explanation for this rule: the relative coordinate of outgoing 1S states cannot absorb the unit of string angular momentum around the axis $\vecr$. In the present formalism an equivalent explanation can be formulated by translating the action of the decay operator into vectors relevant to the initial state:

Isgur, Kokoski and Paton \cite{ikp} observed a more general rule forbidding the decay of $\Pi \uP^\pm$ hybrids to any pair of S-wave mesons with identical spatial wavefunctions, 
\be
\Pi \uP^\pm\nrightarrow \uS+\uS\label{piSS},
\ee
where here neither the initial nor final states are harmonic oscillator eigenstates. The present formalism admits a general proof of this rule, which turns out to be a special case of a more general rule: a state with $\lam$ odd (even) is forbidden to decay to a pair of final states with the same spatial wavefunction coupled to $l_{12}$ even (odd). The spatial wavefunction of any state $\ket {\bX}$ can be expressed as a linear combination of harmonic oscillator basis states
\be
\ket{\bX}=\sum_i c_i\ket{n_il_i}\label{hoexpansion}
\ee
The spatial part of the decay amplitude to any identical pair of states $\ket {\bX}$ is then
\be
\rme{((\bX  \ot\bX  )_{l_{12}}\ot L)_{l_f}}{\grad}{(\lam\ot n'l')_l}	=\sum_{ij}c_ic_jA_{ij}	
\ee
where $A_{ij}$ is the amplitude \rf{hmm1} with $n_il_i$ and $n_jl_j$ in place of $n_1l_1$ and $n_2l_2$. The terms $A_{ij}$ contain Talmi coefficients 
\be
\bk{(n_i'l_i'\ot n_j'l_j')_{l_{12}'}}{(N\lam\ot n''l'')_{l_{12}'}}
\ee
which pick up $(-)^{l''+l_{12}}$ under $i\leftrightarrow j$ and are nonzero only for $(-)^{l_i'+l_j'}=(-)^{\lam+l''}$, using \rf{orbconservation} and \rf{talmisymmetry}. Together with the constraint $(-)^{l_i+l_j}=1$ it is easy to show that $A_{ji}=(-)^{\lam+l_{12}}A_{ij}$ and with the expansion \rf{hoexpansion} above the selection rule follows:
\be
\rme{((\bX  \ot\bX  )_{l_{12}}\ot L)_{l_f}}{\grad}{(\lam\ot n'l')_l}	=0\qquad\textrm{ if }(-)^{l_{12}+\lam}=-1\label{selectionrule}.
\ee
So the decay of any $\Pi$-state to a pair of S-wave mesons is forbidden, as they must be coupled to $l_{12}=0$
\be
\rme{((\bX  \ot\bX  )_{l_{12}=0,2,\ldots}\ot L)_{l_f}}{\grad}{(\Pi\ot n'l')_l}=0\label{piXX}.
\ee
A new result is that the decay of any conventional meson to an identical pair coupled to $l_{12}=1$ is forbidden,
\be
\rme{((\bX  \ot\bX  )_{l_{12}=1,3,\ldots}\ot L)_{l_f}}{\grad}{(\Sigma\ot n'l')_l}=0.\label{sigmaXX}
\ee
Few decay modes are entirely at the mercy of this rule since  the recoupling \rf{separatingspinandspace} maps a final state of definite $j_{12}$ onto a linear combination of  $l_{12}$. The rule does, however, have implications for hybrid production, and so will be drawn upon in subsequent sections.

%%%%%%%%%%%%%%%%%%%%%%%%%%%%%%%%%%%%%%%%%%%%%%%%%%%%%%%%%%%%%%%%%%%%%%%%%%%%%%%%%%%%%%%%%%%%%%%%
\section{Hybrid meson production}
\label{lll}

Consider the most general case in which the final state can also include gluonic excitations with nonzero $\lam_1,\lam_2$:
\be
\lam\to\lam_1+\lam_2.
\ee
Recall that the flux tube quantum numbers $\lam_1,\lam_2$ correspond to operators in the vectors $\vecr_2,\vecr_1$ respectively. Acting with these operators on the final states  $\bra{n_2'l_2'},\bra{n_1'l_1'}$ gives a new pair of harmonic oscillators $\bra{n_2''l_2''},\bra{n_1''l_1''}$. In this way, any amplitude involving gluonic excitations in the final state can be expressed as a linear combination of amplitudes involving only conventional mesons:
\begin{multline}
\rme{
(
	(
		(\lam_1\ot n_1'l_1')_{l_1}
		\ot
		(\lam_2\ot n_2'l_2')_{l_2}
	)_{l_{12}}
	\ot L)_{l_f}}
{\grad}
{(\lam\ot n'l')_l}
=\sum_{n_1''l_1''n_2''l_2''}
\bk{((\lam_1\ot n_1'l_1')_{l_1}\ot (\lam_2\ot n_2'l_2')_{l_2})_{l_{12}}}{(n_1''l_1''\ot n_2''l_2'')_{l_{12}}}
\\\rme{(((n_1''l_1''\ot n_2''l_2'')_{l_{12}})\ot L)_{l_f}}
{\grad}
{(\lam\ot n'l')_l},
\label{brarecoupling}
\end{multline}
where the second term is that calculated in the previous section and summarised in \rf{hmm1}--\rf{hmm3}. The requisite transformation  is found by recoupling the $\lam_1,\lam_2$ operators so as to interchange their positions, and using
\be
\ket{(\lam\ot n'l')_{n''l''}}=\ket{n''l''}\frac{1}{\st{l''}}\rme{n''l''}{\lam}{n'l'}.
\ee
For the general case the recoupling is
\begin{multline}
\bk{((\lam_1\ot n_1'l_1')_{l_1}\ot (\lam_2\ot n_2'l_2')_{l_2})_{l_{12}}}{(n_1''l_1''\ot n_2''l_2'')_{l_{12}}}
\\=(-)^{\lam_1+\lam_2+l_1}\st{l_1l_2}(-)^{l_1''}
\ninej{l_1'}{\lam_1}{l_1}{\lam_2}{l_2'}{l_2}{l_1''}{l_2''}{l_{12}}
\rme{n_1''l_1''}{\lam_2}{n_1'l_1'}\rme{n_2''l_2''}{\lam_1}{n_2'l_2'}.
\label{lamlam}
\end{multline}
If one of the final state mesons is a conventional  meson,
\begin{multline}
\bk{((\lam_1\ot n_1'l_1')_{l_1}\ot (\Sigma \ot n_2'l_2')_{l_2})_{l_{12}}}{(n_1''l_1''\ot n_2''l_2'')_{l_{12}}}
\\=(-)^{l_1+l_{12}+l_2}\st{l_1}
\sixj{l_2''}{\lam_1}{l_2'}{l_1'}{l_{12}}{l_1'}
\rme{n_2''l_2''}{\lam_1}{n_2'l_2'}
\delta_{n_1'n_1''}\delta_{l_1'l_1''},
\label{lamsigma}
\end{multline}
and if both final state mesons are conventional  mesons the recoupling is trivial: 
\be
\bk{((\Sigma\ot n_1'l_1')_{l_1}\ot (\Sigma\ot n_2'l_2')_{l_2})_{l_{12}}}{(n_1''l_1''\ot n_2''l_2'')_{l_{12}}}
=\delta_{n_1'n_1''}\delta_{l_1'l_1''}\delta_{n_2'n_2''}\delta_{l_2'l_2''}.
\label{sigmasigma}
\ee
It follows immediately that these more general amplitudes recover the results of the previous section if both outgoing mesons are of the conventional type. 

The results so obtained are the first calculation of hybrid production amplitudes. Processes such as 
\bea
\Sigma&\to&\Pi+\Sigma\textrm{, and}\\
\Sigma&\to&\Pi+\Pi
\eea
will be discussed elsewhere in the context of $e^+e^-$ $B$-factories \cite{ccdecay}. The next section considers a special case of such processes in which a $1\Pi\uP^-$ hybrid is produced along with a $1\uS$ state. In this case the recoupling \rf{lamsigma} is straightforward, mapping onto a pair of 1P states coupled to $l_{12}=1$:
\be
\rme{(((\Pi\ot 1\uP)_1\ot(\Sigma\ot 1\uS)_0)_1\ot L)_{l_f}}{\grad}{(\lam\ot n'l')_l}
\propto
\rme{ ((1\uP\ot 1\uP)_1\ot L)_{l_f}}{\grad}{(\lam\ot n'l')_l},
\ee
so that, using \rf{selectionrule}, a new selection rule arises
\be
\Sigma\nrightarrow 1\Pi\uP^- + 1\uS.\label{pi1P1S}
\ee
Another way of seeing this is to replace $\Pi$ by the operator $\vecr_2$ and for the harmonic oscillator in $\vecr_1$ that $\ket{1\uP}\propto \vecr_1\ket{1\uS}$, thus
\begin{multline}
\rme{(((\Pi\ot 1\uP)_1\ot(\Sigma\ot 1\uS)_0)_1\ot L)_{l_f}}{\grad}{(\lam\ot n'l')_l}
\\\propto
\rme{(((\vecr_2\ot \vecr_1)_1\ot(1\uS\ot 1\uS)_0)_1\ot L)_{l_f}}{\grad_{\vecr_1}+\grad_{\vecr_2}+ \grad_{\vecr/2}}{(\lam\ot n'l')_l}
%\\\propto
%\rme{(((\vecr_2\ot \vecr_1)_1\ot(1\uS\ot 1\uS)_0)_1\ot L)_{l_f}}{\grad_{\vecr_1}+\grad_{\vecr_2}+ \grad_{\vecr/2}}{(\lam\ot n'l')_l}
\end{multline}
The decay operator then gives two terms, one proportional to 
\be
\bra{((\vecR\ot(1\uS\ot 1\uS)_0)_1\ot L)_{l_f}}\ket{(\lam\ot n'l')_l}
\ee
and another proportional to 
\be
\rme{(((\vecR\ot\vecr)_1\ot(1\uS\ot 1\uS)_0)_1\ot L)_{l_f}}{\vecr+ \grad_{\vecr/2}}{(\lam\ot n'l')_l},
\ee
both of which are zero for an initial $\Sigma$ state on the account of the vector $\vecR$ in the final state.

%%%%%%%%%%%%%%%%%%%%%%%%%%%%%%%%%%%%%%%%%%%%%%%%%%%%%%%%%%%%%%%%%%%%%%%%%%%%%%%%%%%%%%%%%%%%%%%%%%%

\section{The suppressed production of the $1^{-+}$ hybrid}
\label{suppressed}

There has been speculation that the surprisingly large double $c\bar c$ cross sections in $\epem$ $B$-factories could prove a copious source of charmonia hybrids, or likewise the decay of physical $b\bar b$ states. Analogously, it is possible that light quark hybrids could be buried in the wealth of data anticipated on charmonia decays. Such processes will be considered in more detail in REF. \cite{ccdecay}; in this section it will be shown that within the framework developed there the production of the $1\Pi\uP^-$ hybrid along with a conventional 1S meson is forbidden.
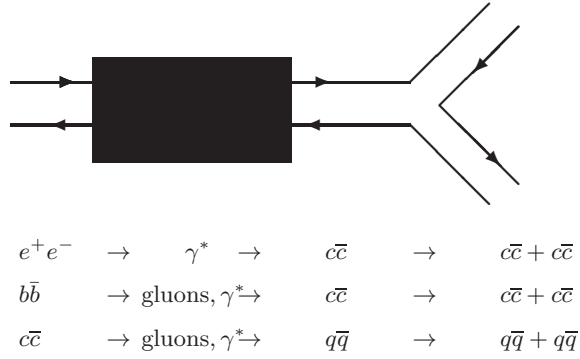
\begin{figure}
%vqpc.tex
\setlength{\unitlength}{0.00023300in}%
\begingroup\makeatletter\ifx\SetFigFont\undefined
% extract first six characters in \fmtname
\def\x#1#2#3#4#5#6#7\relax{\def\x{#1#2#3#4#5#6}}%
\expandafter\x\fmtname xxxxxx\relax \def\y{splain}%
\ifx\x\y   % LaTeX or SliTeX?
\gdef\SetFigFont#1#2#3{%
  \ifnum #1<17\tiny\else \ifnum #1<20\small\else
  \ifnum #1<24\normalsize\else \ifnum #1<29\large\else
  \ifnum #1<34\Large\else \ifnum #1<41\LARGE\else
     \huge\fi\fi\fi\fi\fi\fi
  \csname #3\endcsname}%
\else
\gdef\SetFigFont#1#2#3{\begingroup
  \count@#1\relax \ifnum 25<\count@\count@25\fi
  \def\x{\endgroup\@setsize\SetFigFont{#2pt}}%
  \expandafter\x
    \csname \romannumeral\the\count@ pt\expandafter\endcsname
    \csname @\romannumeral\the\count@ pt\endcsname
  \csname #3\endcsname}%
\fi
\fi\endgroup
\begin{picture}(11713,9663)(1757,-16780)
\thicklines
\put(7951,-10936){\line(-1, 0){3300}}
\put(5101,-10936){\vector(-1, 0){2400}}
\put(5101,-10936){\line(-1, 0){3300}}
\put(10526,-10936){\vector(-1, 0){2000}}
\put(6151,-9961){\vector( 1, 0){3000}}
%\special{ps: gsave 0 0 0 setrgbcolor}
{\linethickness{1.4cm}
\put(3676,- 10586){\line(1,0){4550}}}
%\put(12751,-8161){\vector(-1,-1){1050}}
%\put(10951,-10936){\vector( 1,-1){975}}

%\put(13426,-12286){\vector(-1, 1){975}}
\put(12048,-10908){\vector(  1, -1){975}}

\put(13426,-8686){\vector(-1,-1){1050}}
\put(11626,-10486){\line( 1,-1){1800}}
\put(11626,-10486){\line( 1, 1){1800}}
\put(7651,-10936){\line( 1, 0){3300}}
\put(10951,-10936){\line( 1,-1){1800}}
\put(7651,-9961){\line( 1, 0){3300}}
\put(10951,-9961){\line( 1, 1){1800}}
\put(1801,-9961){\vector( 1, 0){1500}}
\put(5701,-10411){\makebox(6.6667,10.0000){\SetFigFont{10}{12}{rm}.}}
\put(1801,-9961){\vector( 1, 0){1500}}
\put(5701,-10411){\makebox(6.6667,10.0000){\SetFigFont{10}{12}{rm}.}}
\put(2401,-9961){\line( 1, 0){2700}}
\put(1801,-9961){\vector( 1, 0){1500}}
\put(5701,-10411){\makebox(6.6667,10.0000){\SetFigFont{10}{12}{rm}.}}
\put(1801,-9961){\vector( 1, 0){1500}}
\put(5701,-10411){\makebox(6.6667,10.0000){\SetFigFont{10}{12}{rm}.}}
%put(+6200,- 9500){$\bk{\vecR,\vecr}{(\lam\ot n'l')_{lm}}$}
%put(13500,-8500){$\bk{\vecr_2,\vecr_1}{(\lam_1\ot n_1'l_1')_{l_1m_1}}$}
%put(13700,-12500){$\bk{\vecr_1,\vecr_2}{(\lam_2\ot n_2'l_2')_{l_2m_2}}$}
%\put(9001,-12000){$\ket \bA$}
%\put(14001,- 8000){$\ket \bB$}
%\put(14001,-13000){$\ket \bC$}
\put(2001,-14000){$\epem$}\put(4001,-14000){$\to$}\put(5801,-14000){$\gamma^*$}\put(7001,-14000){$\to$}
\put(9001,-14000){$\cc$}\put(11001,-14000){$\to$}\put(13001,-14000){$\cc+\cc$}
\put(2001,-15000){$b\bar b$}\put(4001,-15000){$\to$}\put(4801,-15000){$\textrm{gluons},\gamma^*$}\put(7001,-15000){$\to$}
\put(9001,-15000){$\cc$}\put(11001,-15000){$\to$}\put(13001,-15000){$\cc+\cc$}
\put(2001,-16000){$\cc $}\put(4001,-16000){$\to$}\put(4801,-16000){$\textrm{gluons},\gamma^*$}\put(7001,-16000){$\to$}
\put(9001,-16000){$\qq$}\put(11001,-16000){$\to$}\put(13001,-16000){$\qq+\qq$}
\end{picture}
\caption{Some possible sources of hybrid states in $\epem$ $B$-factories.}
\label{blackbox}
\end{figure}

The processes under consideration share common features as depicted in FIG. \ref{blackbox}: annihilation into a virtual photon or gluons and the creation of a virtual quark pair which then decays via pair creation. In the approach of REF. \cite{ccdecay} only the second part of the process is calculated -- the strong decay of a virtual quark-antiquark pair  into the final state pair. In a non-relativistic potential picture, the pair that emerge from the black box can be any admixture of orbital, radial or gluonic excitations consistent with the initial state quantum numbers, and $a$ $priori$ the   heirachy of mixing angles between such states is unknown. Arguments given in REF. \cite{ccdecay} suggest that gluonically excited modes are suppressed relative to $\Sigma$ modes. As an example consider the $1^{--}$ channel, whose intermediate states could be  admixtures of $\ket {n^3\Sigma \uS_1}$, $\ket {n^3\Sigma \uD_1}$ and $\ket{n^1\Pi\uP^-_1}$.  Any significant $\Pi$ admixture would be in conflict with known prevalence of  experimental modes such as
\bea
\epem&\to&\eta_c\psi,\\
\psi &\to&\rho \pi ,\\
\psi&\to& b_1\pi,	
\eea
the first two on account of the selection rule \rf{piSS} and last because of the spin-singlet selection rule, a zero in the 9-$j$ coefficient in \rf{spinpart} for $s=s_1=s_2=0$. This suppression may be expected because inside the black box of FIG. \ref{blackbox} will be factors sensitive to the wavefunction at the origin, suggesting $\ket {n^3\Sigma \uS_1}$ states will dominate. For channels with quantum numbers other than $1^{--}$ and $0^{-+}$ both the $\Sigma$ and glue-excited states have nodes at the origin so equivalent  arguments do not apply,  but there is phenomenological evidence suggesting the $\Sigma$ states dominate. For $\chi_2$ decay, for instance, the intermediate state could be populated by $\ket{n^3\Sigma\uP_2}$ and $\ket{n^1\Pi\uD^+_2}$ but the prevalence of modes such as
\be
\chi_2\to\phi\phi,\omega\omega,\pi\pi,\eta\eta
\ee
would argue against the importance of the latter, for here the selection rule \rf{piXX} is expected to be exact. Deferring further arguments to REF. \cite{ccdecay}, suppose that 
the gluonically-excited modes can be neglected, in which case  the spatial wavefunction of the pair emerging from the black box can be any admixture of $\Sigma$ states consistent with the quantum numbers of the initial state.
%:
%\bea
%\ket \bA&=&\sum_i \alpha_i\ket{(\Sigma\ot n_il_i)_{l_i}}\\
	%&=&\sum_i \alpha_i\ket{n_il_i}.
%\eea
All such processes as depicted in FIG. \ref{blackbox} are then be subject to the selection rule \rf{pi1P1S},
\be
\Sigma\nrightarrow 1\Pi\uP^- + 1\uS,
\ee
valid in the limit of equal width wavefunctions for the final states but independent of the radial wavefunction of the intial state. This has implications for the lightest hybrid states $\ket{1^1\Pi\uP^-_1}$ with $1^{--}$ quantum numbers and the $\ket{1^3\Pi\uP^-_{0,1,2}}$ with $(0,1,2)^{-+}$ quantum numbers. In charmonia decay, for instance, some channels whose quantum numbers are ripe for the production of the exotic $1^{-+}$ should be forbidden in this limit:
\bea
\psi&\nrightarrow& \pi_1\rho\textrm{ in P-, F-wave, and}\label{psitopi1rho}\\
\chi_1&\nrightarrow& \pi_1\pi\textrm{ in S-, D-wave.}\label{chitopi1pi}
\eea
Strong observation of such modes in the $\pi_1(1400),\pi_1(1600)$ channels would argue against a hybrid interpretation for them, although the different spatial wavefunctions of the $\pi$ and $\pi_1$ may lead to significant breaking of the selection rule \cite{ccdecay}. For charmonia states the rule would be expected to hold more strongly: this would argue against a hybrid  interpretation for the 
$X(3940)$ observed in
\be
\epem\to J/\psi X(3940)\label{x3940},
\ee
if it is found to have $J^{-+}$ quantum numbers, but not neccessarily if it has $1^{++}$. A hybrid interpretation is thought to be untenable in any case, on account of the low mass. On the other hand if the $1^{--}$ $Y(4260)$ is a $\ket{1^1\Pi\uP^-_1}$ hybrid then it should have $(0,1,2)^{-+}$ partners $\ket{1^3\Pi\uP^-_{0,1,2}}$ in the same mass region; because of the selection rule, the absence of such peaks in the data is understandable. The selection rule may help discriminate between hybrid and other exotic or non-exotic interpretations of heavy or light-quark states, include those with non-exotic quantum numbers. For instance, searching for forbidden modes such as 
\be
\Upsilon,\epem\nrightarrow Y(4260)\eta_c
\ee
may help discriminate intepretations of the $Y(4260)$.

%%%%%%%%%%%%%%%%%%%%%%%%%%%%%%%%%%%%%%%%%%%%%%%%%%%%%%%%%%%%%%%%%%%%%%%%%%%%%%%%%%%%%%%%%%%%%%%%%%%%%%%%%%%%%%%
\section{conclusion}

The Talmi-Moshinsky transformation has been applied to the strong decay problem and several new results derived. The method allows easy derivation of previous results  for conventional mesons  and extends upon those to encompass decays involving any orbital and radial excitations. By factorising the string degrees of freedom amplitudes for decays involving gluonic excitations are a simple generalisation of those involving only conventional mesons. Many of the amplitudes so obtained have not previously been considered; their final polynomial forms will be collated elsewhere and applied to conventional and hybrid meson production in charmonia decays \cite{ccdecay}. 

Within this formalism the well-known selection rule against the decay of a hybrid to a pair of identical 1S mesons has a natural explanation. Using the Talmi-Moshinsky transformation it is possible to generalise this rule to the case of identical wavefunction that need not be harmonic oscillators, in agreement with numerical calculations with realistic wavefunctions. There appears a new class of selection rules which have implications for the production of $1^{-+}$ exotic hybrids and may help discriminate interpretations of states with non-exotic quantum numbers.  Owing to its being purely geometrical in nature, the Talmi-Moshinksy transformation may find application to other quark model processes with the geometry of FIG. \ref{geometry}. There are generalisations of the Talmi-Moshinksy transformation for systems described by more than two Jacobi coordinates and these may be useful in, for instance, meson rescattering amplitudes or processes involving baryons.

The flux-tube model naturally suggests a decay operator with pair creation in spin 1 and provides a parameter-free scale for processes involving hybrids relative to those involving only conventional mesons. Lattice QCD is now able to calculate strong decays and these two features of the model appear to be confirmed \cite{ftlattice}. Comparing to the lattice allows models to be tested and refined; at the same time models can shed light on the dynamics of the underlying QCD of the lattice. Further comparisons of the type \cite{ftlattice} can be made easily using the results of the present work. The formalism developed here could also be applied  to decay amplitudes via the $\tso$ operator and to more general processes allowing for spin- or orbital-flips at the pair creation vertex.

Generalisations of the operator approach employed here may be of use to the problem of rescattering between hybrid and conventional meson states, using the ideas of \cite{greenpaton}. Likewise it may be possible to employ similar techniques to model flux-tube de-excitation decay modes, predicted in lattice calculations to be significant for hybrid states \cite{mmp}; this could help in the interpretation of the Y(4260) state.

The recoupling \rf{brarecoupling} that relates the spatial wavefunction of outgoing hybrid states to those of conventional mesons can also be performed in the $\ket{(s\ot nl)_j}$ basis. In this way, final states involving hybrids can be expressed as a superposition of conventional meson states and their production amplitudes relative to conventional mesons is independent of the initial state; this is applied to hybrid pair production elsewhere \cite{ccdecay}.

That the wavefunctions in one set of Jacobi coordinates translate into a finite number of wavefunctions in the other is a feature peculiar to harmonic oscillator wavefunctions. The calculations presented here could be used as a basis for calculations with more realistic wavefunctions: the coefficients in the translation from these to the harmonic oscillator basis are well-known. It is straightforward to generalise these calculations to allow for different wavefunction width in the initial state or to include a localised pair creation region with spherical symmetry: these would appear only in the final integrals \rf{radialR} and \rf{radialr}. Those integrals could also be performed numerically to incorporate the cigar-shaped pair creation region of \cite{ki}. 

%It is interesting to note that a heavier  $1^{-+}$ state can be formed from $\ket {\Pi^-1\uD}$ as in equation \rf{pim1D}; production of such a state from the above processes would not be forbidden.
%%%%%%%%%%%%%%%%%%%%%%%%%%%%%%%%%%%%%%%%%%%%%%%%%%%%%%%%%%%%%%%%%%%%%%%%%%%%%%%%%%%%%%%%%%%%%%%%%%%%%%%%%%%%%%%

\vspace{1cm}

This work is supported by a Clarendon Fund Bursary. For many useful discussions the author is indebted to Jack Paton, Jo Dudek and particularly Frank Close.

\bibliographystyle{unsrt}
\bibliography{tjb}

\end{document}